\documentclass[aps,prb,twocolumn,showpacs]{revtex4-1}
\usepackage{bm}
\usepackage{graphicx}
\usepackage{amsmath,amsthm,amssymb}
\usepackage{dsfont}
\usepackage{soul}
\usepackage{hyperref}

\usepackage{color}

\begin{document}

\title{Nonmonotonic quantum phase gathering in curved spintronic circuits}

\author{Eusebio J. Rodr\'iguez}
\email{erfernandez@us.es}
\affiliation{Departamento de F\'isica Aplicada II, Universidad de Sevilla, E-41012 Sevilla, Spain}

\author{Diego Frustaglia}
\email{frustaglia@us.es}
\affiliation{Departamento de F\'isica Aplicada II, Universidad de Sevilla, E-41012 Sevilla, Spain}

\date{\today}

\begin {abstract}
Spin carriers propagating along quantum circuits gather quantum spin phases depending on the circuit's size, shape, and spin-orbit coupling (SOC) strength. These phases typically grow monotonically with the SOC strength, as found in Rashba quantum wires and rings. In this work we show that the spin-phase gathering can be engineered by geometric means, viz. by the geometric curvature of the circuits, to be non-monotonic. We demonstrate this peculiar property by using one-dimensional polygonal models where flat segments alternate with highly curved vertices. The complex interplay between dynamic and geometric spin-phase components--- triggered by a series of emergent spin degeneracy points--- leads to bounded, global spin phases. Moreover, we show that the particulars of the spin-phase gathering have observable consequences in the Aharonov-Casher conductance of Rashba loops, a connection that passed unnoticed in previous works.   
\end{abstract}

\maketitle

\section{Introduction}

Spin dynamics can be determinant for electronic transport in mesoscopic conductors. \cite{review-spintronics2020} Carriers developing spin-dependent phases experience quantum interference effects molding properties such as the conductance. Coherent spin-phase contributions can be sufficiently strong to reverse the magnetoconductance response of low-dimensional systems, as in the case of weak (anti)localization.\cite{B82,MMKN05}
A prominent source of spin phases in two-dimensional electron gases is spin-orbit coupling (SOC), responsible for the working principles of spin-field-effect transistors \cite{DD90,MK01,KKECHJ09,CHSSPCFGFBJRC15} and Aharonov-Casher (AC) interferometers, \cite{AC84,MS92,NMT99,FR04,MPV04,BKSN06,KTHSHDSBBM06,GLIERW07} among others.\cite{MKNFD15,BL15} Moreover, when SOC is combined with Zeeman fields and superconducting proximity effects, spin carriers can develop more exotic quantum states of topological nature such as the celebrated (and elusive) Majorana modes.\cite{MZFPBK12} 

The electrical modulation of the Rashba SOC \cite{BR84} in two-dimensional electron gases confined in semiconductor heterostructures\cite{NATE97} has facilitated the realization of spin interferometers \cite{BKSN06,KTHSHDSBBM06,GLIERW07} based on the AC effect.\cite{AC84} This is an electrical effect on a particle carrying a quantum magnetic moment, which is nothing but the electromagnetic dual of the Aharonov-Bohm (AB) effect \cite{AB59} (a magnetic effect on an electrically charged quantum particle). The role played by dynamic and geometric spin phases in the conductance of AC interferometers (specially in semiconductor-based mesoscopic rings) has been studied intensively over the last decade.\cite{R12,NTKKN12,NFSRN13,SRBFN18,NRBFSN18,FN20} Moreover, further studies on polygonal AC interferometers \cite{BFG05,vVKN06,KSN06,QYCSDLSYLFLJYL11,WSRBFN19,HBFB21}  have demonstrated that the conductance is quite sensitive to the geometric shape of the conducting channels (specifically, to their curvature\cite{O15,YGOC16,YGBFOC20,SSAJ21}) due to the development of strongly non-adiabatic spin dynamics.\cite{PFR03} This has significant consequences on the response to external fields and the generation of topological spin phases.\cite{WSRBFN19} Still, some questions remain open in this regard as the interplay between dynamic and geometric spin-phase gathering and their distinct contributions.    

Here, we address these questions by studying the development of dynamic and geometric phases in spin carriers propagating through one-dimensional model loops of polygonal shape subject to Rashba SOC. By these means, we find that the spin-phase gathering in Rashba polygons is strongly non-monotonic, in manifest contrast to what observed in Rashba rings. Our results show that this is a direct consequence of spin degeneracies emerging from the non-adiabatic spin dynamics triggered by field discontinuities at the polygon vertices. These features lead to a series of remarkable effects such as, e.g., the bounding of global AC spin phases, the possibility of purely geometric spin-phase gathering (due to vanishing dynamic spin phases), and the development of geometric spin-phase plateaus that allow the independent control of dynamic spin phases (complementary to previous findings regarding the independent control of geometric spin phases in rings\cite{NFSRN13}).

We point out that one-dimensional models for spin-carrier transport in mesoscopic interferometers have been used in the past with success. \cite{LGC90,S92,AL-G93,FR04,BFG05,NFSRN13,HBFB21,WSRBFN19} In particular, models similar to the one employed here have demonstrated to be well suited to experiments with arrays of interferometric loops  where only one single (quasi-one-dimensional) orbital mode appears to contribute to quantum interference due to the decoherence experienced by relatively slow propagating higher modes.\cite{NFSRN13,WSRBFN19}

The paper is organized as follows. In Sec. \ref{model} we introduce a one-dimensional model for conducting Rashba polygons. Our results on non-monotonic spin-phase gathering are presented in Sec. \ref{spin-phases}. In Sec. \ref{conductance} we discuss the consequences of the spin-phase gathering process in the conductance of Rashba loops. Sec. \ref{conclusions} is devoted to closing comments and conclusions. We also include a series of appendices with additional discussions on spin dynamics in Rashba loops and a semiclassical approach to the quantum conductance.

\section{Model}
\label{model}

\begin{figure}
\includegraphics[width=\columnwidth]{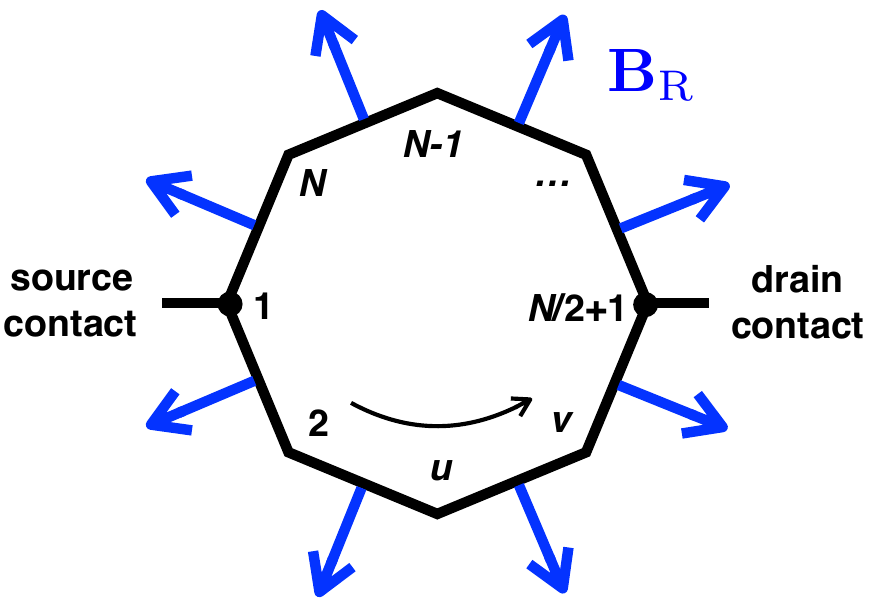}
\caption{One-dimensional Rashba polygon model. The effective Rashba field ${\textbf B}_{\text R}$ correspond to CCW propagating spin carriers.}
\label{fig-1}
\end{figure}

Consider a regular polygon of perimeter $P$ lying on the $xy$-plane consisting of $N$ conducting segments of length $L = P/N$ connecting vertices $u$ and $v$ and oriented along directions $\hat{\boldsymbol{\gamma}}_{vu}$ (from $u$ to $v$), see Fig. \ref{fig-1}. The spin-carrier dynamics along each wire segment is determined by the Hamiltonian\cite{BFG05}
\begin{equation}
\label{Hvu}
H_{vu}=\frac{p_\ell^2}{2m}+\frac{\alpha_{\text R}}{\hbar} p_\ell (\hat{\boldsymbol{\gamma}}_{vu} \times \hat{\bf z})\cdot \boldsymbol{\sigma},
\end{equation}
with $p_\ell=-i\hbar\partial_\ell$ the linear momentum of the spin-carriers, 
$\ell$ the linear coordinate along the wire, $m$ the carrier's mass, $\boldsymbol{\sigma}$ the vector of Pauli matrices,
and $\alpha_{\text R}$ the Rashba SOC strength (which can be controlled in experiments by electrical means \cite{NATE97}). The second term in Eq. (\ref{Hvu}) represents an effective in-plane magnetic field ${\bold B}_{\text R}= (2\alpha_{\text R}/\hbar g \mu_{\text B})p_\ell(\hat{\boldsymbol{\gamma}}_{vu} \times \hat{\bf z})$ coupled to the itinerant spins, with $g$ the $g$-factor and $\mu_{\text B}$ the Bohr magneton. Notice that ${\bold B}_{\text R}$ is momentum-dependent and normal to $\hat{\boldsymbol{\gamma}}_{vu}$, inverting its sign for counterpropagating carriers. This means that $H_{vu}$ preserves time-reversal symmetry.

By completing squares in Eq. (\ref{Hvu}) we find
\begin{equation}
\label{Hvu-2}
H_{vu}= \frac{1}{2m}\left(p_\ell+ A_{vu}\right)^2-\frac{\hbar^2}{2m}k_{\text R}^2,
\end{equation}
where $A_{vu}=\hbar k_{\text R}(\hat{\boldsymbol{\gamma}}_{vu} \times \hat{\bf z})\cdot \boldsymbol{\sigma}$ and $k_{\text R}= \alpha_{\text R}m/\hbar^2=\pi/\lambda_{\text R}$, with $\lambda_{\text R}$ the spin-precession length. We notice in Eq. (\ref{Hvu-2}) that the SOC term of Eq. (\ref{Hvu}) has turned into a gauge field $A_{vu}$ (playing the role of a spin-dependent vector potential) and a uniform spin-independent energy offset (which can be disregarded). This means that the solutions of the Schroedinger equation $H_{vu}|\psi\rangle=E|\psi\rangle$ are plane waves such that initial spin-carrier states $|\psi(0)\rangle$ injected in vertex $u$ propagate along the segment towards vertex $v$ as 
\begin{equation}
\label{wf}
|\psi(\ell)\rangle=e^{-ik_{\text F}\ell}e^{-i k_{\text R} \ell (\hat{\boldsymbol{\gamma}}_{vu} \times \hat{\bf z})\cdot \boldsymbol{\sigma}}|\psi(0)\rangle,
\end{equation}
with $k_{\text F}$ the Fermi wavenumber. The first prefactor on the rhs of Eq. (\ref{wf}) corresponds to the Abelian $U(1)$ kinetic phase of the carrier associated to the charge's dynamics, while the second prefactor represents the non-Abelian $SU(2)$ spin phase due to spin precession.\cite{HBFB21} 

Equations (\ref{Hvu})-(\ref{wf}) indicate that spin carriers propagating in polygonal Rashba loops undergo a series of effective-field discontinuities at the vertices due to the abrupt changes of $\hat{\boldsymbol{\gamma}}_{vu}$, defining a textured ${\bold B}_{\text R}$ with significant consequences for the spin dynamics (eventually leading to the development of complex paths in the Bloch sphere after the spin carriers complete a round trip along the polygonal circuit, as demonstrated in previous theoretical\cite{BFG05,vVKN06,HBFB21} and experimental works\cite{KSN06,QYCSDLSYLFLJYL11,WSRBFN19}). Still, we notice that actual semiconductor-based circuits of polygonal shape may present rounded vertices that would soften the discontinuities and, eventually, its effects on the carriers' spin evolution. However, one can show (see Appendix \ref{appendix-1}) that this would require very large SOC fields outside the actual range of interest and beyond current experimental reach in mesoscopic systems. More precisely, it would entail the spin-precession length $\lambda_{\text R}$ to be much smaller that the effective size of the vertices for a spin to notice their rounded shape.\cite{footnote-1} As a consequence, for all practical purposes we find that the modelling of the polygons' vertices as point-like discontinuities is fully justified.  

Moreover, one finds that ring-shaped loops can be modelled as polygons by taking the limit $N \gg 1$, provided that $L \ll \lambda_{\text R}$. In this limit, an effective radial ${\bold B}_{\text R}$ emerges and the polygonal shape remains unnoticed\cite{footnote-1} by the itinerant spins as shown in previous works\cite{BFG05,HBFB21} (see also Appendix \ref{appendix-1}).  

From Eq. (\ref{wf}) we find that the propagation of a spin carrier from $u$ to $v$ is fully determined by the phases $k_{\text F} L+ k_{\text R} L (\hat{\boldsymbol{\gamma}}_{vu} \times \hat{\bf z})\cdot \boldsymbol{\sigma}$. In particular, the spin evolution along a full segment is given by the momentum-independent spin rotation operator
\begin{equation}
\label{Rvu}
R_{vu}=\exp[-i k_{\text R} L (\hat{\boldsymbol{\gamma}}_{vu} \times \hat{\bf z})\cdot \boldsymbol{\sigma}],
\end{equation}
with $R_{vu}^\dagger=R_{uv}$ due to time-reversal symmetry. Equation (\ref{Rvu}) is the building block to describe the spin evolution of a carrier propagating in a polygonal loop. By labelling the vertices from $1$ to $N$, we find that the spin evolution along counterclockwise (CCW) and clockwise (CW) propagating paths is given by the unitary operators
\begin{eqnarray}
\label{UN+}
U_+(N)&=&R_{1N}...R_{32} \ R_{21}, \\
\label{UN-}
U_-(N)&=&R_{12}...R_{N-1,N} \ R_{N1},
\end{eqnarray}
respectively, with $U_-(N)=U_+^\dagger(N)$. The eigenvalue equation  
\begin{equation}
\label{eigeneqn}
U_\pm(N)|\chi_s\rangle=\exp[\pm i\phi_s] |\chi_s\rangle\\
\end{equation}
defines the global AC spin phase $\pm \phi_s$ gathered by a carrier after a CCW/CW round trip propagating from the initial state $|\chi_s\rangle$ to a final state $\exp[\pm i\phi_s] |\chi_s\rangle$ defined at the initial vertex 1, with $s=\uparrow,\downarrow$ and $\langle \chi_\uparrow|\chi_\downarrow \rangle=0$. In regular polygons, symmetry dictates that the spin quantization axis $\hat{\bold n}_s=\langle \chi_s|\boldsymbol{\sigma}|\chi_s\rangle$ is contained within the plane normal to the polygon that bisects the vertex's angle. This hampers the full alignment of $\hat{\bold n}_s$ and ${\bold B}_{\text R}$, compelling the state $|\chi_s\rangle$ to propagate along the corresponding segment by precessing around the local ${\bold B}_{\text R}$. This repeats identically for every vertex and segment. It is only in the limiting case of ring-shaped loops ($N \gg 1$, $L \ll \lambda_{\text R}$) that the regime of adiabatic spin dynamics\cite{FR04} can be approached for $\lambda_{\text R} \ll P$ and the spin eigenmodes tend to align with the local ${\bold B}_{\text R}$ (see Appendix \ref{appendix-1}). The different cases are illustrated by the insets in Fig. \ref{fig-2} representing the corresponding \emph{spin textures}, i.e., the circulation path described by the local spin states in the Bloch sphere as the carriers propagate completing a round trip.

The global AC spin phase $\phi_s=\phi_{\text d}^s+\phi_{\text g}^s$ splits into dynamic ($\phi_{\text d}^s$) and geometric ($\phi_{\text g}^s$) phase components.\cite{AA87} The dynamic spin phase represents the expectation value of the spin Hamiltonian over the propagating spin modes in a CCW round trip (i.e., the projection of the spin texture on the effective-field texture). Since the contributions to the spin phases along each segment are identical due to symmetry, the dynamic phase reduces to $\phi_{\text d}^s = -k_{\text R} P (\hat{\boldsymbol{\gamma}}_{21} \times \hat{\bf z})\cdot \hat{\bold n}_s$. The geometric spin phase $\phi_{\text g}^s=-\Omega_s/2$, instead, is proportional to the solid angle $\Omega_s$ subtended by the spin texture of CCW propagating modes. This geometric spin phase, also referred to as the Ahronov-Anandan phase,\cite{AA87} converges to a Berry phase\cite{B84} only in the limit of adiabatic spin dynamics (disfavoured in Rashba polygons). The global and dynamical spin phases, $\phi_s$ and $\phi_{\text d}^s$, are obtained easily by solving Eq. (\ref{eigeneqn}). An explicit calculation of the geometric spin phase $\phi_{\text g}^s$ is sometimes difficult, but unnecessary in our case: it can be readily obtained from the other two as their difference, $\phi_{\text g}^s=\phi_s-\phi_{\text d}^s$. 

%

A similar analysis applies to CW propagating spin carriers by replacing $\phi_s$ with $-\phi_s$, as seen from Eq. (\ref{eigeneqn}).

\section{Results}
\label{results}

We start by calculating the global ($\phi_s$), dynamic ($\phi_{\text d}^s$), and geometric ($\phi_{\text g}^s$) spin phases gathered by carriers in polygonal loops with an even $N$. The results exhibit a rich spin dynamics as compared to ring-shaped loops, with several possibilities for spin-phase manipulation. We then show how these features reflect in the transport properties of polygonal loops. For these aims, we introduce a dimensionless $k_{\text R} P$ to quantify the Rashba SOC strength. This corresponds to the spin phase gathered by a spin carrier propagating along a straight quantum wire of length $P$, used here as a blank for evaluating spin phases in curved circuits. Transport experiments in mesoscopic rings\cite{BKSN06,KTHSHDSBBM06,GLIERW07,NTKKN12,NFSRN13,NRBFSN18} and, especially,  squares\cite{KSN06,QYCSDLSYLFLJYL11,WSRBFN19} with perimeters of few micrometers show that $k_{\text R} P$ can be electrically modulated in a wide range covering several multiples of $2\pi$ relevant to our discussion.

\subsection{Spin phases}
\label{spin-phases}

\begin{figure}
\includegraphics[width=\columnwidth]{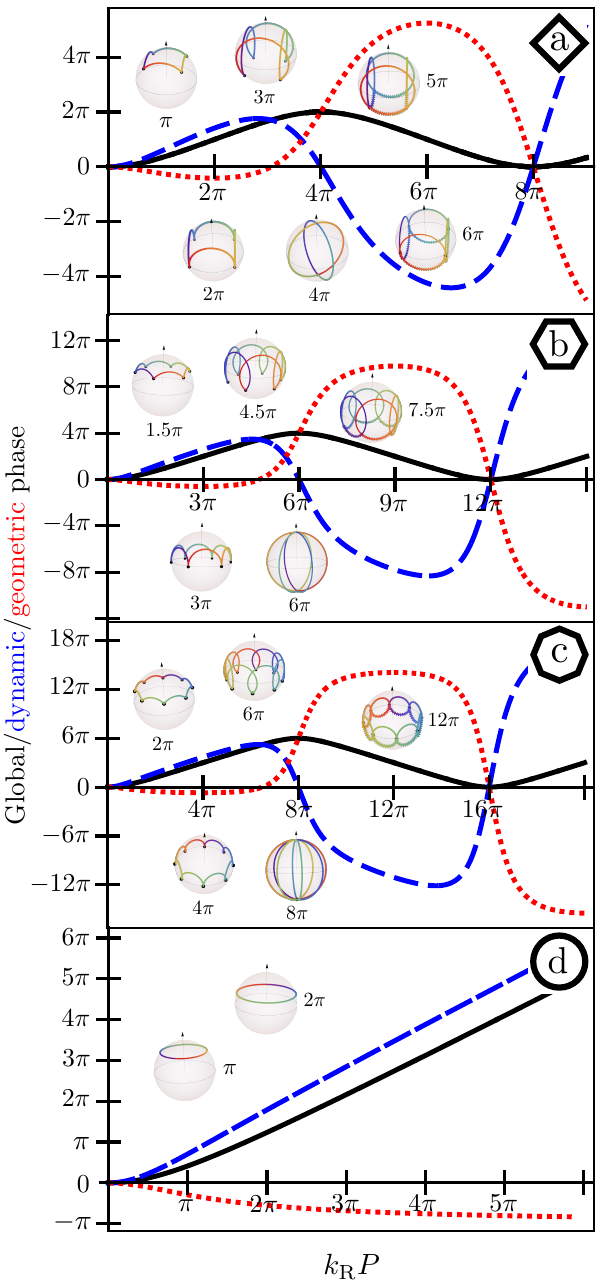}
\caption{Global (black solid lines), dynamic (blue dashed lines) and geometric (red dotted lines) phases of propagating spin modes in Rashba squares (a), hexagons (b), octagons (c), and rings (d) as a function of the Rashba SOC strength $k_{\text R}P$. The points of vanishing dynamic phase correspond to spin degeneracies. The Bloch-sphere insets depict the spin textures of propagating modes for different values of $k_{\text R}P$ (the color scale shows the circulation direction--- from red to violet--- starting a round trip from vertex 1; the solid dots indicate the local spin states at the vertices). The complex response of the spin phases and textures for polygons contrast with the monotonic response for rings.}
\label{fig-2}
\end{figure}

\begin{figure}
\includegraphics[width=\columnwidth]{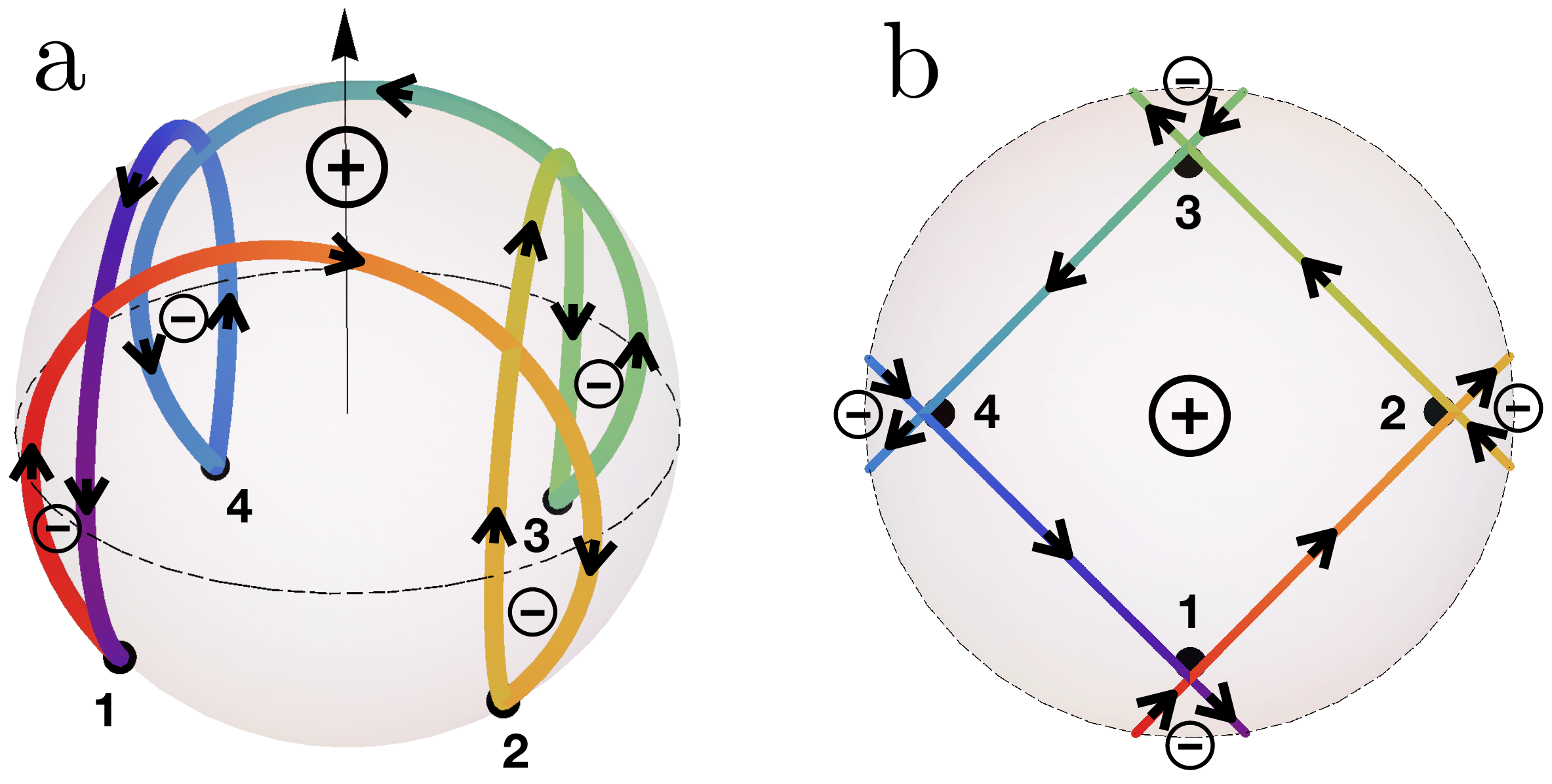}
\caption{(a) Bloch sphere showing the spin texture of a propagating mode with vanishing geometric phase in a Rashba square corresponding to $k_{\text R}P=3\pi$ , see Fig. \ref{fig-2}(a). The arrows indicate the circulation of the local spin states as the carrier propagate along the perimeter and the color scale represents the position. The numbers label the local spin states at the vertices. Positive and negative signs indicate the contributions from each section of the spin texture to its solid angle. (b) Azimuthal projection of the spin texture.
}
\label{fig-3}
\end{figure}

For the sake of simplicity we focus our attention on CCW propagating spin-up ($s=\uparrow$) carriers. We define the spin-up species as the branch for which $|\chi_\uparrow \rangle \rightarrow |\uparrow \rangle_z$ as the Rashba SOC vanishes ($k_{\text R}P \rightarrow 0$), where $|\uparrow \rangle_z$ is the eigenstate of  $\sigma_z$ with positive eigenvalue. Henceforth, we drop the spin label from the global ($\phi$), dynamic ($\phi_{\text d}$), and geometric ($\phi_{\text g}$) spin phases.

In Fig. \ref{fig-2} we plot the discriminated spin phases as a function of the Rashba SOC strength $k_{\text R}P$ for (a) squares, (b) hexagons, (c) octagons, and (d) rings (modelled by using $N=60$, $10 L \le \lambda_{\text R}$). It stands out the strongly non-monotonic response of the spin phases in polygons, contrasting with a purely monotonic behavior in the ring's case. We observe that the global spin phase $\phi$ (black solid lines) for polygons is \emph{bounded}, $0 \le \phi \le(N-2)\pi$, oscillating with periodicity $2N\pi$ as a function of $k_{\text R}P$. This unexpected bounding is the consequence of a singular interplay between dynamic and geometric phase components in polygons, $\phi_{\text d}$ (blue dashed lines) and $\phi_{\text g}$ (red dotted lines), which oscillate counterphase with an increasing amplitude. 

In Figs. \ref{fig-2}(a), \ref{fig-2}(b) and \ref{fig-2}(c), the oscillating dynamic spin phase $\phi_{\text d}$ (blue dashed lines) reveals aspects of an intricate spin evolution (see the corresponding spin textures in the Bloch-sphere insets). Its sign is indicative of the spin-state projection along the local Rashba field ${\bf B}_{\text R}$ (negative projection for positive $\phi_{\text d}$, and viceversa). Its growing amplitude is an expected consequence of the increasing Rashba-field strength, even for partial spin-state/field alignment. The most interesting feature is the vanishing of $\phi_{\text d}$ at $k_{\text R}P= n N\pi$ ($n$ integer), representing spin degeneracy points. This condition is equivalent to $L=n\lambda_{\text R}$, corresponding to an integer number of spin precession along each side of the polygon. At the degeneracy points the spinors $|\chi_s \rangle$ at vertex 1 quantized along the $z$-axis, such that spins propagate along the polygons perimeter by precessing within a plane perpendicular to the local Rashba field ${\bf B}_{\text R}$. These states are illustrated in Figs.  \ref{fig-2}(a), \ref{fig-2}(b) and \ref{fig-2}(c) by the insets corresponding to $k_{\text R}P=4\pi, 6\pi$, and $8\pi$, respectively. 

As for the geometric spin phase $\phi_{\text g}$ (red dotted lines), its sign indicates the dominating direction of circulation of the spin states in the Bloch sphere while its magnitude grows with the subtended solid angle. Notice that a vanishing $\phi_{\text g}$, except for the case of $k_{\text R}P=0$, is not a signal of pinned spin states but, instead, an indicator of complex spin textures where sections with different circulation directions in the Bloch sphere contribute with partial solid angles of opposite sign that cancel each other. An example in Rashba squares, Fig. \ref{fig-2}(a), takes place at $k_{\text R}P=3\pi$: the corresponding spin texture is shown with further detail in Fig. \ref{fig-3} where positive and negative contributions to the solid angle are identified. The periodic development of opposite contributions of this kind lead to the oscillating response of $\phi_{\text g}$ in polygons shown in Figs. \ref{fig-2}(a), \ref{fig-2}(b) and \ref{fig-2}(c). Two particular features stand out here: (i) $\phi_{\text g}$ tends to develop plateaus in the vicinity of its extremes. This tendency is more pronounced as the number of sides $N$ increases. Interestingly, within a geometric-phase plateau the global phase $\phi$ presents a linear evolution as a function of $k_{\text R}P$ with origin in the dynamic component $\phi_{\text d}$. This shows the possibility of an independent control of the dynamic spin-phase component $\phi_{\text d}$ in Rashba polygons, complementary to the purely geometric spin phase manipulation achieved in Rashba rings by introducing weak inplane Zeeman fields.\cite{NFSRN13} (ii) At the degeneracy points ($\phi_{\text d}=0$), the global phase reduces to a purely geometric phase of magnitude $\phi_{\text g}=(N-2)\pi$. For an even $N$, this multiple of $2\pi$ corresponds to the geometric phase associated to the solid angle of $N/2-1$ full spheres. This means that between consecutive degeneracies the geometric phase undergoes an integer number of windings $w_{\text g}=\phi_{\text g}/2\pi=N/2-1$, contributing to characterize the spin dynamics in topological terms.

The above description differs from what observed in Fig. \ref{fig-2}(d) for Rashba rings. There, both the global and dynamic spin phases, $\phi$ (black solid line) and $\phi_{\text d}$ (blue dashed line), grow monotonically with $k_{\text R}P$. This growing becomes linear as the spin dynamics turns adiabatic for large $k_{\text R}P$. The adiabatic regime is particularly well illustrated by the evolution of the geometric spin phase $\phi_{\text g}$ (red dotted line), approaching $-\pi$ for large $k_{\text R}P$ as expected for spin states aligned with the radial Rashba field. In this limit, the spin states orbit the equator of the Bloch sphere by subtending a solid angle corresponding to half sphere. Figure \ref{fig-2}(d) represents an effective decoupling between dynamic and geometric spin phases in rings, which is absent in polygons due to the emergence of degeneracy points. 

We stress that for producing Fig. \ref{fig-2}(d) we modelled the Rashba ring by using a polygon with relatively large $N=60$ while keeping the spin precession length much larger than the polygon's sides ($k_{\text R}P \le 6\pi$, equivalent to $10 L \le \lambda_{\text R}$ in this case). This model would fail as $k_{\text R}P$ approaches $60\pi$ and $\lambda_{\text R} \rightarrow L$, where the first spin degeneracy emerges.

Moreover, a closer look at Fig. \ref{fig-2} actually shows that in the regime of weak Rashba SOC strengths $k_{\text R}P \ll N\pi$ (namely, far from first degeneracy point and equivalent to $L \ll \lambda_{\text R}$) the spin phases in any polygon mimic the response observed in rings. Deep in that weak-field limit, Rashba polygons and rings look very similar from the point of view of the gathered spin phases. See Appendix \ref{appendix-1} for a discussion.

All these aspects are addressed analytically in Appendix \ref{appendix-2} for the particular case of Rashba squares. 

We further notice that recent magnetotransport experiments\cite{WSRBFN19} in Rashba squares with $P \approx 2.8 \mu m$ demonstrate that Rashba SOC strengths can be electrically tuned in a window $4\pi \le k_{\text R}P \le 6\pi$, a vicinity of the first degeneracy point where the polygonal shape manifests optimally.

\subsection{Conductance}
\label{conductance}

Here we show that the spin-phase characteristics discussed in the previous Sec. \ref{spin-phases} have observable consequences in the conductance of Rashba polygons. To this aim, we consider Rashba polygons with $N$ (even) sides symmetrically coupled to source and drain contact leads as shown in Fig. \ref{fig-1}. We adopt the Landauer-B\"uttiker formulation\cite{BILP85} at zero temperature by identifying the linear conductance $G$ with the quantum transmission $T$ (in units of the quantum of conductance $e^2/h$): $G=(e^2/h) T$, with $T=\sum_{mn}|t_{mn}|^2$ and $t_{mn}$ the quantum transmission amplitude from the incoming mode $n$ at the source contact lead to the outgoing mode $m$ at the drain contact lead. In our one-dimensional model we have one single orbital mode and two spin modes, such that $0 \le T \le2$. Moreover, the unitarity of the scattering matrix imposes $T+R=2$, where $R=\sum_{mn}|r_{mn}|^2$ is the quantum reflection with $r_{mn}$ the corresponding amplitudes for incoming and outgoing modes $n$ and $m$ at the source contact lead. 

A realistic modelling of the experimental conditions, as those corresponding to two-dimensional Rashba loop arranges,\cite{NTKKN12,NFSRN13,WSRBFN19} requires to take into account the effect of disorder and/or sample averaging. By following a semiclassical approach for loops strongly coupled to the contact leads (see Appendix \ref{appendix-3}) we distinguish two different situations: 

(i) Systems preserving a two-fold reflection symmetry along the axis connecting the contact leads. In this case, the basic traits of the quantum conductance are captured by the expression
\begin{equation}
\label{G1}
G_1=\frac{e^2}{h}(1+\cos \phi),
\end{equation}
with $\phi$ the global spin phase defined in Sec. \ref{spin-phases}. The resulting interference pattern, oscillating as a function of $k_{\text R}P$ (through $\phi$), is the AC effect in Rashba loops.\cite{AC84,NMT99,FR04}

(ii) Systems with broken geometrical symmetries. In this case, the quantum conductance is best described by the expression
\begin{equation}
\label{G2}
G_2=\frac{e^2}{h}(1-\cos 2\phi).
\end{equation}
Equation (\ref{G2}) captures the pairing of time-reversed orbital path emerging in disordered systems. In the absence of Rashba SOC ($\phi=0$), this pairing leads to a minimum in the quantum conductance due to the constructive interference of backscattered carriers, an effect known as weak localization. For sufficiently strong Rashba SOC (e.g., $\phi=\pi/2$), this effect is reversed by destructive interference of backscattered spin carriers maximizing the quantum conductance, leading to the so-called weak antilocalization. This pattern oscillates with a frequency two times larger than the one observed for $G_1$ in Eq. (\ref{G1}). This frequency doubling shares its origin with the Al'tshuler-Aronov-Spivak oscillations\cite{AAS81} observed in magnetoconductance experiments with disordered rings and squares.\cite{NTKKN12,NFSRN13,WSRBFN19}  Hence, the relevance of either $G_1$ or $G_2$ for a given implementation can be decided independently in the laboratories by performing complementary magnetoconduntance measurements and observing the periodicity of the oscillations in units of the magnetic flux quantum $\phi_0=hc/e$. 

\begin{figure}
\includegraphics[width=\columnwidth]{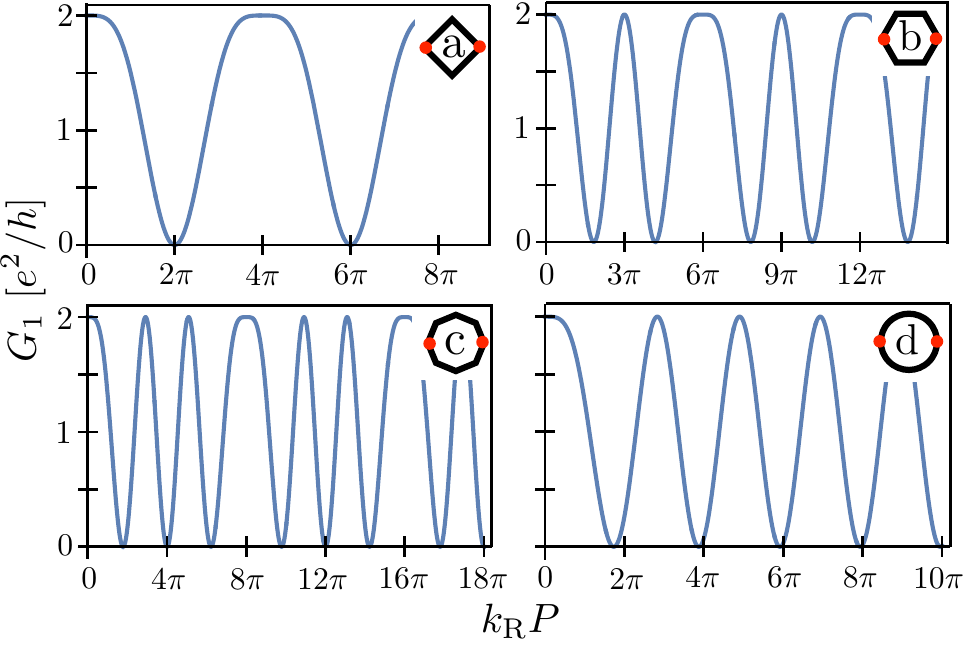}
\caption{Conductance $G_1$ of Eq. (\ref{G1}) as a function of the Rashba coupling strength $k_{\text R}P$ corresponding to two-fold symmetric (a) squares, (b) hexagons, (c) octagons, and (d) rings. Insets: Our semiclassical model considers spin carriers propagating along paths starting in one red spot (left) and ending in other one (right).}
\label{fig-4}
\end{figure}

\begin{figure}
\includegraphics[width=\columnwidth]{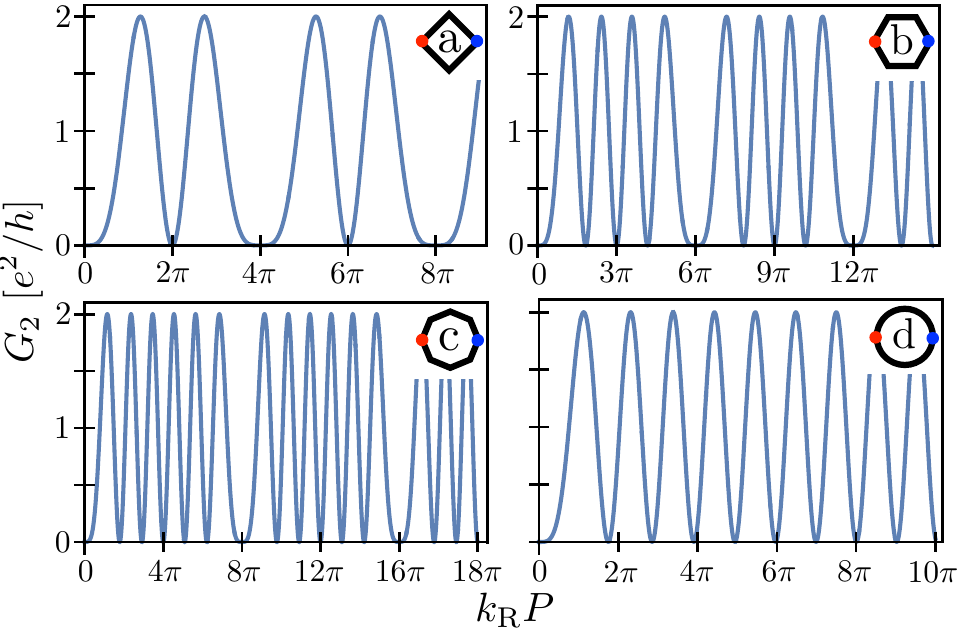}
\caption{Conductance $G_2$ of Eq. (\ref{G2}) as a function of the Rashba coupling strength $k_{\text R}P$ corresponding to disordered (a) squares, (b) hexagons, (c) octagons, and (d) rings. Insets: Our semiclassical model considers spin carriers propagating along paths starting and ending at the red spots (left). Notice the frequency doubling with respect to the results of Fig. \ref{fig-4} due to time-reversed path interference (dual of the magnetoconductance Al'tshuler-Aronov-Spivak oscillations). The minima for $k_{\text R}P=0$ is due to weak localization.}
\label{fig-5}
\end{figure}

Figures \ref{fig-4} and \ref{fig-5} illustrate the results of Eqs. (\ref{G1}) and (\ref{G2}), respectively, by plotting $G_1$ and $G_2$ as a function of $k_{\text R}P$. Two particular features stand out there: (i) The extremes of the global phase $\phi$--- related to degeneracy points due to vanishing dynamic phases, as shown in Fig. \ref{fig-2}--- manifest as conductance plateaus (see Appendix \ref{appendix-2} for relevant analytic expressions in Rashba squares). 
(ii) Away from the degeneracy points, in the regime where the global phase $\phi$ responds linearly to $k_{\text R}P$, the conductance displays rapid AC oscillations dominated by the dynamic spin phase. These features explain the presence of two different frequencies contributing to the AC conductance oscillations in polygons: a lower frequency given by the periodicity of the global phase and a higher frequency determined by large phase-gathering rates between global phase extremes. One consequence is that the lower frequency contribution dominates in squares while it is absent in rings. Such contributing frequencies have been previously identified\cite{BFG05} and discussed\cite{HBFB21} in terms of length scales, i.e., the perimeter $P$ and the segments' length $L=P/N$ measured in units of the spin-precession length $\lambda_{\text R}$. However, the non-monotonicity of the gathered spin phases and the presence of emergent spin degeneracies passed unnoticed in those discussions. 

\section{Conclusions}
\label{conclusions}

We have demonstrated that spin carriers propagating in circuits of polygonal shape gather spin phases in a non-monotonic fashion as a function of the Rashba SOC strength. This peculiar behaviour is triggered by the inhomogeneities of the geometric curvature along the polygonal perimeter--- where flat segments alternate with highly curved vertices--- introducing effective Rashba field discontinuities acting as scattering centers for spin. As a consequence, a periodic series of spin degeneracy points emerge. This contrasts with the case of ring circuits of constant curvature, which present a monotonous phase gathering and a complete absence of spin degeneracies. We find that the global spin phases 
oscillate with a period determined by the emergent degeneracies. Moreover, dynamic and geometric spin-phase components present a rich pattern allowing the independent control of dynamic phases over geometric-phase plateaus (complementary to previous findings on the independent control of geometric spin phases in Rashba rings\cite{NFSRN13}). 

We have also shown that the non-monotonicity of the spin-phase gathering has definite consequences in AC conductance oscillations as the presence of plateaus in the vicinity of spin degeneracies and the participation of two different frequencies. These frequencies were identified in previous works\cite{BFG05,HBFB21} but interpreted in terms of the different length scales present in the system, overlooking the particulars of the spin-phase gathering and the existence of spin degeneracies.

Our findings offer original ways to engineer electronic spin phases at the mesoscopic scale by geometric means. The identification of emergent degeneracies suggest that non-abelian geometric spin phases may also apply. The role played by commensurability effects in the development of emergent degeneracies and phase gathering remains an open question, motivating future investigations on irregular polygons. The relevance shown by the geometric curvature in the reported findings also suggest that prospective studies of spin dynamics in non-euclidean surfaces are in order.

Finally, non-monotonic phase gathering and quantum phase bounding arise as physical phenomena worthy of being singled out, as they could be identified and exploited in other two-level quantum systems subject to complex dynamics. 

\begin{acknowledgments}
This work was supported by the Spanish Ministerio de Ciencia, Innovaci\'on y Universidades through Project No. FIS2017-86478-P and by the Andalusian Government through the PAIDI 2020 initiative with Project No. P20-00548. We acknowledge useful discussions with J. Rahn and J.P. Baltan\'as.
\end{acknowledgments}


\appendix

\section{Adiabatic condition in Rashba rings and polygons}
\label{appendix-1}

\begin{figure}
\includegraphics[width=\columnwidth]{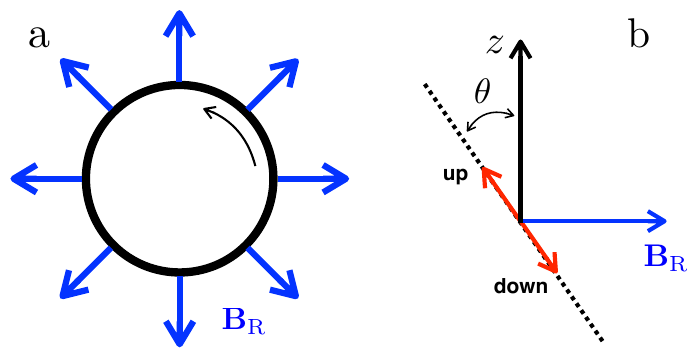}
\caption{
One-dimensional Rashba ring. (a) Travelling spin carriers experience an effective (momentum-dependent) radial magnetic field ${\bold B}_{\text R}$ (pointing outwards for CCW travellers and inwards for CW ones). (b) Spin eigenstates quantize along an axis with tilt angle $\theta$ such that $\tan \theta =Q_{\text R}$. This axis coincides with ${\bold B}_{\text R}$ only in the adiabatic limit $Q_{\text R} \gg 1$.
}
\label{fig-6}
\end{figure}

The spin-carrier dynamics in a one-dimensional Rashba ring of radius $r_0$ lying on the $xy$-plane (see Fig. \ref{fig-6}) is determined by the Hamiltonian\cite{MMK02,FR04,FN20}
\begin{eqnarray}
\label{H0a}
H_0&=&-\frac{\hbar\omega_0}{2}\partial_\varphi^2+\frac{\hbar\omega_{\text R}}{2}\sigma_r(-i\partial_\varphi)-i\frac{\hbar\omega_{\text R}}{4}\sigma_\varphi\\
\label{H0b}
&=&\frac{\hbar\omega_0}{2} \left(-i\partial_\varphi +\frac{Q_{\text R}}{2}\sigma_r\right)^2-\frac{\hbar\omega_0}{8}Q_{\text R}^2,
\end{eqnarray}
with $\varphi$ the polar angle parametrizing the ring's perimeter and Pauli matrices $\sigma_r=\cos\varphi \ \sigma_x+\sin\varphi \ \sigma_y$ and $\sigma_\varphi=-\sin\varphi \ \sigma_x+\cos\varphi \ \sigma_y$. Moreover, we have defined the characteristic kinetic and Rashba SOC frequencies $\omega_0=\hbar/mr_0^2$ and $\omega_{\text R}=2\alpha_{\text R}/\hbar r_0$, respectively, and their quotient $Q_{\text R}=\omega_{\text R}/\omega_0$. The second term in Eq. (\ref{H0a}) represents an effective (momentum-dependent) radial magnetic field (${\bold B}_{\text R}$ in Fig. \ref{fig-6}) coupled to the itinerant spins, while the last term (negligible in the semiclassical limit of large momenta) is introduced to guarantee the Hermiticity of the SOC.\cite{MMK02} 

From Eq. (\ref{H0b}) one finds spin eigenstates of the form\cite{FR04,FN20}
\begin{eqnarray}
\label{psi-up}
|\psi_{l,\uparrow} \rangle&=&\exp(i l \varphi) \left( 
\begin{array}{c}
\cos\theta/2\\ 
-e^{i\varphi} \sin\theta/2
\end{array}
\right),\\
\label{psi-down}
|\psi_{l,\downarrow} \rangle&=&\exp(i l \varphi) \left( 
\begin{array}{c}
\sin\theta/2\\ 
e^{i\varphi} \cos\theta/2
\end{array}
\right),
\end{eqnarray}
where the integer $l=\pm k r_0$ is the angular momentum of the carriers (with $\pm$ for CCW/CW motion and $k$ the wavenumber). The eigenstates (\ref{psi-up}) and (\ref{psi-down}) define spin textures of conic shape in the Bloch sphere as a function of $\varphi$, as those depicted by the insets in Fig. \ref{fig-2}(d). These spin textures are fully determined by the parameter $Q_{\text R}$ since $\sin \theta = Q_{\text R}/\sqrt{1+Q_{\text R}^2}$, $\cos \theta = 1/\sqrt{1+Q_{\text R}^2}$, and $\tan \theta = Q_{\text R}$, with $\theta$ the angle between the local spin quantization axis and the $z$-axis. The corresponding dynamical and geometric phases read $\phi_{\text d}^s=s\pi Q_{\text R} \sin \theta$ and $\phi_{\text g}^s=-\pi (1- s \cos \theta)$.\cite{FN20} These phases are non-vanishing (except for $Q_{\text R}=0$) and experience a monotonous growing with $Q_{\text R}$. 

We identify the regime of adiabatic spin dynamics, where the spin quantization axis is locally aligned with the radial field ${\bold B}_{\text R}$, as the limiting case $Q_{\text R} \gg 1$ (i.e., $\theta \rightarrow \pi/2$). In this adiabatic limit, the spin texture describes a path along the equator of the Bloch sphere by subtending a solid angle $2\pi$ and gathering a geometric (Berry) phase equal to $-\pi$. In the opposite limit $Q_{\text R} \ll 1$  (i.e., $\theta \rightarrow 0$) one finds that the spin eigenstates tend to stay pinned at the poles of the Bloch sphere along the ring's perimeter. This means that the spin carriers are practically unperturbed by the radial field texture ${\bold B}_{\text R}$. The crossover regime of finite $Q_{\text R}$ is generally referred to as the non-adiabatic regime.

It results useful to rewrite the parameter $Q_{\text R}$ as the ratio of two length scales, viz. $Q_{\text R}=2\pi r_0/\lambda_{\text R}$. This means that the adiabatic regime (where the spin carriers capture all the details of the radial field texture) requires the spin precession length, $\lambda_{\text R}$, to be much smaller than the circumference of the ring, $2\pi r_0$. In the opposite limit, $2\pi r_0 \ll \lambda_{\text R}$, the radial field texture passes unnoticed to the spin carriers. \cite{footnote-1} 

\begin{figure}
\includegraphics[width=\columnwidth]{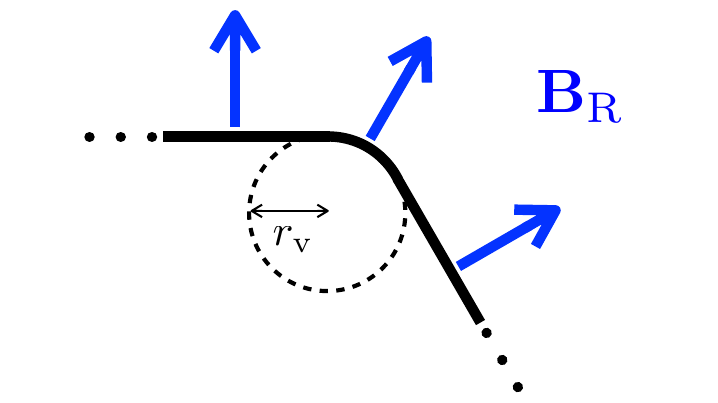}
\caption{
Detail of a polygonal one-dimensional Rashba loop with rounded vertices, modelled as arcs of circumference with radius $r_{\text v}$. The effective field ${\bold B}_{\text R}$ is uniform along each segment and radial along the arcs. The spin carriers experience field discontinuities at the vertices in the regime $\lambda_{\text R} \gg 2\pi r_{\text v}$.
}
\label{fig-7}
\end{figure}

Consider now a Rashba loop of polygonal shape of perimeter $P$ with $N$ rounded vertices modelled as arcs of circumference with radius $r_{\text v} \ll P/N$ (see Fig. \ref{fig-7}). 
The effective inplane Rashba SOC field ${\bold B}_{\text R}$ experienced by the itinerant spin carriers is uniform along the segments and radial along the arcs. The conditions for adiabatic spin dynamics are determined by the regions  with maximal field-texture inhomogeneity, corresponding to maximal geometric curvature in this case.\cite{YGBFOC20} This means that the adiabatic condition in a polygon with rounded vertices coincides with that of a small Rashba ring of radius $r_{\text v}$, viz. $\lambda_{\text R} \ll 2\pi r_{\text v}$. This would require relatively large Rashba SOC strengths, far beyond the regime $\lambda_{\text R} \approx P/N$ where the first spin degeneracy points emerge (corresponding to $k_{\text R}P=N\pi$ in Fig. \ref{fig-2}). This means that the vertices can be safely treated as point-like discontinuities for moderate Rashba strengths covering the first degeneracy points. 

Finally, we identify a regime where adiabatic spin dynamics in Rashba rings can be modelled by using polygons. The polygonal nature of Rashba circuits manifests in the spin dynamics when the Rashba SOC strength is such that $\lambda_{\text R} \approx P/N$ and the first degeneracy point arises. Hence, by working in the regime of much weaker Rashba SOC strengths, viz. $\lambda_{\text R} \gg P/N$, the effects of the polygonal shape can be minimized. Moreover, as discussed above, the adiabatic condition in a Rashba ring of radius $r_0$ is set by $\lambda_{\text R}\ll 2\pi r_0$. We then find that the regime of adiabatic spin dynamics can be approached by using polygons with $P=2\pi r_0$ provided that $P/N \ll \lambda_{\text R} \ll P$, which is possible for large $N$ thanks to a length-scale separation. This is confirmed in our simulations of Fig. \ref{fig-2}(d) as well as in previous works.\cite{BFG05,HBFB21}

\section{Spin dynamics in Rashba square loops}
\label{appendix-2}

\begin{figure}
\includegraphics[width=\columnwidth]{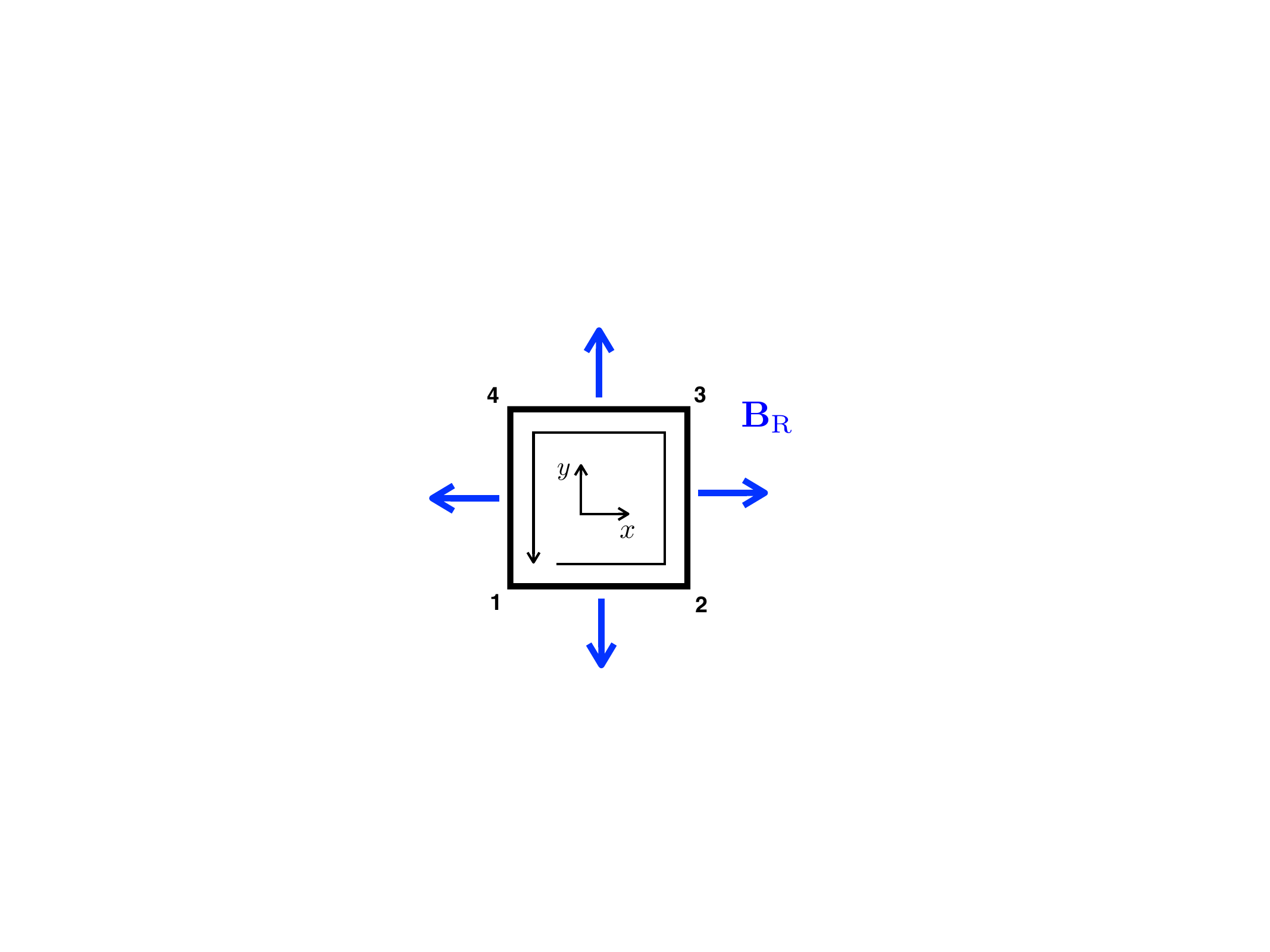}
\caption{
One-dimensional Rashba square model. The effective field ${\bold B}_{\text R}$ corresponde to CCW propagating carriers.
}
\label{fig-8}
\end{figure}

The spin dynamics of CCW propagating carrier in a Rashba square of perimeter $P=4L$ is determined by the unitary operator introduced in Eq. (\ref{UN+}) with $N=4$
\begin{equation}
\label{U4+}
U_+(4)=R_{14}R_{43}R_{32}R_{21},
\end{equation}
with
\begin{eqnarray}
\label{R21-4}
R_{21}&=&\exp[i k_{\text R} L \sigma_y],\\
R_{32}&=&\exp[-i k_{\text R} L \sigma_x],\\
R_{43}&=&\exp[-i k_{\text R} L \sigma_y],\\
\label{R14-4}
R_{14}&=&\exp[i k_{\text R} L \sigma_x],
\end{eqnarray}
where we use the reference system depicted in Fig. \ref{fig-8}. As any $SU(2)$ operator, Eq. (\ref{U4+}) can be written as
\begin{equation}
\label{U4+bis}
U_+(4)=\exp[i\phi \ \hat{\bold n}\cdot \boldsymbol{\sigma}]=\cos\phi \ \mathbb{I}+i \sin\phi \ \hat{\bold n}\cdot \boldsymbol{\sigma},
\end{equation}
satisfying the eigenvalue equation
\begin{equation}
\label{U4+eigen}
U_+(4)|\chi_s\rangle=\exp[is\phi]|\chi_s\rangle,
\end{equation}
with spinors 
\begin{eqnarray}
\label{chi-up}
|\chi_{\uparrow} \rangle&=& \left( 
\begin{array}{c}
\cos\theta_0/2\\ 
e^{i\varphi_0} \sin\theta_0/2
\end{array}
\right),\\
\label{chi-down}
|\chi_{\downarrow} \rangle&=&\left( 
\begin{array}{c}
\sin\theta_0/2\\ 
-e^{i\varphi_0} \cos\theta_0/2
\end{array}
\right),
\end{eqnarray}
defined at the initial vertex $1$ with quantization axis $\hat{\bold n}= \sin\theta_0 \cos\varphi_0 \ \hat{\bold x}+\sin\theta_0 \sin\varphi_0 \ \hat{\bold y}+\cos\theta_0 \ \hat{\bold z}$. The $\phi$ in (\ref{U4+bis}) and (\ref{U4+eigen}) is nothing but the global AC spin phase gathered by the up species in a round trip. Fron Eqs. (\ref{U4+})-(\ref{U4+bis}) we find 
\begin{eqnarray}
\label{cos-phi}
\cos \phi &=& 1-2\sin^4(k_{\text R}L)\\
\label{sin-phi}
\sin \phi &=& 2\cos(k_{\text R}L)\sin^2(k_{\text R}L) \sqrt{1+\sin^2(k_{\text R}L)} ,\\
\label{cos-theta}
\cos \theta_0 &=&\frac{\cos(k_{\text R}L)}{\sqrt{1+\sin^2(k_{\text R}L)}},\\
\label{sin-theta}
\sin \theta_0 &=&\frac{\sqrt{2}\sin(k_{\text R}L)}{\sqrt{1+\sin^2(k_{\text R}L)}},\\
\varphi_0&=& \pi/4.
\end{eqnarray}
This shows that the spinors (\ref{chi-up}) and (\ref{chi-down}) are quantized within the plane normal to the square bisecting vertex 1, as expected from symmetry. This repeats in all vertices, which can be confirmed by propagating (\ref{chi-up}) and (\ref{chi-down}) with (\ref{R21-4})-(\ref{R14-4}). During propagation along the segments, the spinors precess with constant projection around the local field ${\bold B}_{\text R}$. Notice that the projection $\hat{\bold n} \cdot {\bold B}_{\text R} \propto \sin\varphi_0 \sin\theta_0$ is always partial, meaning that the spinors never align with ${\bold B}_{\text R}$ and the regime of adiabatic spin dynamics does not exist here. The dynamical spin phase $\phi_{\text d}$ is proportional to that projection and equal to 
\begin{equation}
\label{phi-d-4}
\phi_{\text d}=k_{\text R}P\frac{\sin(k_{\text R}L)}{\sqrt{1+\sin^2(k_{\text R}L)}},
\end{equation}
where we exploited the fact that equal phases are gathered along each segment. Equation (\ref{phi-d-4}) shows that $\phi_{\text d}$ oscillates with growing amplitude as a function of $k_{\text R}L$, vanishing at $k_{\text R}L=n\pi$ with integer $n$ (equivalent to $k_{\text R}P=n4\pi$ or $L=n\lambda_{\text R}$). This vanishing points correspond to spin degeneracies. Notice from (\ref{cos-theta}) and (\ref{sin-theta}) that the spinors (\ref{chi-up}) and (\ref{chi-down}) quantize along the $z$-axis at the degeneracy points, meaning that they precess within a plane perpendicular to the local ${\bold B}_{\text R}$ during propagation by describing meridian lines in the Bloch sphere. See blue dashed line and insets in Fig. \ref{fig-2}a for an illustration.

From (\ref{cos-phi}) and (\ref{sin-phi}) we can see that the global spin phase $\phi$ is non-monotonic and bounded (see black solid line in Fig. \ref{fig-2}a). Notice that the plateaus exhibited by the conductances (\ref{G1}) and (\ref{G2}) around the degeneracy points--- illustrated in Figs. \ref{fig-4}a and \ref{fig-5}a--- can be understood directly from Eq. (\ref{cos-phi}). 

As for the geometric phase $\phi_{\text g}=\phi-\phi_{\text d}$, its response is more complex and better appreciated in Fig. \ref{fig-2}a (dotted red line). This geometric phase is a purely Aharonov-Anandan one,\cite{AA87} since the adiabatic limit necessary for the development of Berry phases\cite{B84} can not be reached in Rashba squares (unless rounded vertices and sufficiently large Rashba SOC strengths are considered, see Appendix \ref{appendix-1}).

Finally, we evaluate the response of the spin phases in the limit of weak SOC strengths $k_{\text R}L \ll \pi$ (far from the first degeneracy point) and compare it with the corresponding result for rings of the same perimeter $P$. In this limit, Rashba squares show
\begin{eqnarray}
\phi&=&\frac{1}{8}(k_{\text R}P)^2,\\
\phi_{\text d}&=&\frac{1}{4}(k_{\text R}P)^2,\\
\phi_{\text g}&=&-\frac{1}{8}(k_{\text R}P)^2.
\end{eqnarray}
From Appendix \ref{appendix-1} we find that, in the same limit, Rashba rings show
\begin{eqnarray}
\phi&=&\frac{1}{2\pi}(k_{\text R}P)^2,\\
\phi_{\text d}&=&\frac{1}{\pi}(k_{\text R}P)^2,\\
\phi_{\text g}&=&-\frac{1}{2\pi}(k_{\text R}P)^2.
\end{eqnarray}
We observe that the spin phases gathered by Rashba squares and rings show a similar quadratic response in the weak-field limit. They only differ in a geometrical prefactor of the same order, which would turn it difficult (though not impossible) to distinguish one from the other in this regime. See Figs. \ref{fig-2}a and \ref{fig-2}d, Figs. \ref{fig-4}a and \ref{fig-4}d, and Figs. \ref{fig-5}a and \ref{fig-5}d near the origin for a comparison. 

We finally notice that recent experiments\cite{WSRBFN19} in mesoscopic Rashba squares demonstrate that the SOC strength  can be tuned by electrical means in the vicinity of the first degeneracy point ($k_{\text R}P \sim 4\pi$) where the polygonal shape manifests optimally.

\section{Semiclassical conductance}
\label{appendix-3}

The Landauer-B\"uttiker formulation\cite{BILP85} identifies the two-contact linear conductance $G$ with the quantum transmission and reflection as
\begin{equation}
\label{LB}
G=\frac{e^2}{h} \text{tr}\left[\mathbb{T}\mathbb{T}^\dagger\right]=\frac{e^2}{h} \text{tr}\left[\mathbb{I}-\mathbb{R}\mathbb{R}^\dagger\right],
\end{equation}
with $\mathbb{T}=[t_{mn}]$ and $\mathbb{R}=[r_{mn}]$, where $t_{mn}$ and $r_{mn}$ are the quantum transmission and reflection amplitudes from incoming ($n$) to outgoing ($m$) modes. The trace of $\mathbb{I}$ equals the number of available incoming modes. A semiclassical model \cite{reviews-semiclassics} of $G$ for one-dimensional Rashba loops can be developed whenever the carriers wavelength is much smaller than the system size and the spin splitting is much smaller than the kinetic energy (so that the spin dynamics does not alter the orbital one),\cite{FG93} in agreement with mesoscopic experimental conditions.\cite{NTKKN12,NFSRN13,WSRBFN19} In this way, by following a path-integral approach and taking the semiclassical limit,\cite{LG92} the quantum transmission and reflection amplitudes can be expressed as
\begin{eqnarray}
\label{tmn}
t_{mn}&=&\sum_\Gamma a_\Gamma e^{ik_{\text F}L_\Gamma} \langle m | U_\Gamma | n\rangle, \\
\label{rmn}
r_{mn}&=& \sum_\Gamma b_\Gamma e^{ik_{\text F}L_\Gamma} \langle m | U_\Gamma | n\rangle,
\end{eqnarray}
namely, as a sum of phase contributions over different classical paths $\Gamma$ of length $L_\Gamma$ taking the spin carriers from entrance to exit leads with different statistical weights $a_\Gamma$ and $b_\Gamma$, eventually leading to quantum interference. Within this picture, charge and spin contributions are clearly differentiated. The charge contributes with the orbital phase $\exp[ik_{\text F}L_\Gamma]$. As for the spin, carriers entering the system with spin $n$ can leave it with spin $m$ according to the path-dependent spin evolution operator $U_\Gamma$, which is determined by the particular fields experienced by the spin carriers along the classical path. As for the quantum transmission and reflection, they consists of probability terms of the form
\begin{eqnarray}
\label{Tmn}\\
\nonumber
|t_{mn}|^2&=&\sum_{\Gamma,\Gamma'} a_\Gamma a_{\Gamma'}^*e^{ik_{\text F}(L_\Gamma-L_{\Gamma'})} \langle m | U_\Gamma | n\rangle  \langle m | U_{\Gamma'} | n\rangle^*, \\
\nonumber
|r_{mn}|^2&=&\sum_{\Gamma,\Gamma'} b_\Gamma b_{\Gamma'}^*e^{ik_{\text F}(L_\Gamma-L_{\Gamma'})}  \langle m | U_\Gamma | n\rangle  \langle m | U_{\Gamma'} | n\rangle^*.\\
\label{Rmn}
\end{eqnarray}
For a realistic modelling of the experimental conditions, the effects of disorder and/or sample averaging need to be taken into account. This means that the sums in (\ref{Tmn}) and (\ref{Rmn}) need to run over different configurations including classical path fluctuations. Moreover, an average over a small energy window around the Fermi energy can be also implemented to take into account the effects of finite (though low) temperatures. Due to the presence of the orbital-phase factors $\exp[ik_{\text F}(L_\Gamma-L_{\Gamma'})]$, the averaging procedure shows that the only surviving terms in (\ref{Tmn}) and (\ref{Rmn}) are those corresponding to pairs of paths $\{\Gamma,\Gamma'\}$ with the same geometric length, $L_\Gamma = L_{\Gamma'}$. Other contributions simply average out due to rapid oscillations of the orbital-phase factors. However, identifying these pairs of paths contributing to the transmission in (\ref{Tmn}) is generally difficult unless two-fold reflection symmetry along the axis connecting the contact leads is preserved. Otherwise, it results most convenient to resort to the quantum reflection (\ref{Rmn}) by taking advantage of unitarity. Moreover, when the conducting loops are well coupled to the leads, the carriers tend to escape after a few windings. In this case, it has been shown that the most relevant features of the conductance are fully captured by considering only the shortest paths.\cite{FR04,NFSRN13,WSRBFN19} 

By assuming two-fold symmetric configurations and well-coupled leads, the conductance can be calculated from the simplified transmission amplitudes
\begin{equation}
\label{tmn0}
t_{mn}=\frac{1}{2} \langle m |U_+(N/2+1)+U_-(N/2+1)| n\rangle,
\end{equation}
with
\begin{eqnarray}
\label{UN2+}
U_+(N/2+1)&=&R_{N/2+1,N/2}...R_{32} \ R_{21}, \\
\label{UN2-}
U_-(N/2+1)&=&R_{N/2+1,N/2+2}...R_{N-1,N} \ R_{N1},
\end{eqnarray}
and $R_{vu}$ defined in Eq. (\ref{Rvu}). Notice that we have dropped a phase prefactor $\exp[ik_{\text F}P/2]$ from Eq. (\ref{tmn0}) corresponding to CCW/CW orbital paths of length $P/2$, irrelevant to the transmission (\ref{Tmn}). By using time-reversal symmetry ($R_{vu}^\dagger=R_{uv}$) and working in the eigenbasis of $U_\pm(N)$, Eq. (\ref{eigeneqn}), we find from Eq. (\ref{LB}) that the conductance within this approximation takes the form\cite{G1G2}
\begin{equation}
G_1=\frac{e^2}{h}(1+\cos \phi),
\end{equation}
with $\phi$ the global spin phase gathered by the carriers in a round trip. 

However, in most experimental situations the two-fold symmetry can not be assumed. This general case can be modelled by resorting to the reflection probabilities (\ref{Rmn}) after noticing that, for any backscattering path $\Gamma$, there exists another path $\tilde{\Gamma}$ with exactly the same length that follows the trajectory defined by $\Gamma$ but in opposite direction. Namely, $\Gamma$ and $\tilde{\Gamma}$ are time-reversed paths. By considering well-coupled leads, we find that they correspond to CCW/CW single-winding paths of length $P$. The corresponding reflection amplitudes take the form
\begin{equation}
\label{rmn0}
r_{mn}=\frac{1}{2} \langle m |U_+(N)+U_-(N)| n\rangle,
\end{equation}
where we have dropped a phase prefactor $\exp[ik_{\text F}P]$, irrelevant to the reflection (\ref{Rmn}). By turning to time-reversal symmetry and the eigenbasis of $U_\pm(N)$, in this case we find a conductance\cite{G1G2}
\begin{equation}
G_2=\frac{e^2}{h}(1-\cos 2\phi),
\end{equation}
where the minus sign and the factor 2 in the argument are the consequences of the time-reversed path pairing.

See Appendix \ref{appendix-2} for relevant analytic expression concerning $\phi$ in Rashba squares. 


\begin{equation}
\nonumber
\text{tr}\left[\mathbb{T}\mathbb{T}^\dagger\right]=\frac{1}{2} \text{tr}\left[ \mathbb{I}+
\left( \begin{array}{cc}
e^{i\phi} & 0 \\
0 & e^{-i\phi}
\end{array} \right)
\right] = 1+\cos\phi,
\end{equation}
with 
\begin{equation}
\nonumber
\mathbb{T}=\frac{1}{2} \left[U_+(N/2+1)+U_-(N/2+1)\right].
\end{equation}
Similarly, we find
\begin{equation}
\nonumber
\text{tr}\left[\mathbb{R}\mathbb{R}^\dagger\right]=\frac{1}{2} \text{tr}\left[ \mathbb{I}+
\left( \begin{array}{cc}
e^{i2\phi} & 0 \\
0 & e^{-i2\phi}
\end{array} \right)
\right] = 1+\cos2\phi,
\end{equation}
with 
\begin{equation}
\nonumber
\mathbb{R}=\frac{1}{2} \left[U_+(N)+U_-(N)\right].
\end{equation}


\begin{thebibliography}{99}
\bibitem{footnote-1}
Recalling optical microscopy, where objects smaller than the wavelength of light can not be properly distinguished.
\bibitem{G1G2}
By using the properties of the trace we find that
\end{thebibliography}
\end{document}